\documentclass{ws-procs9x6}
\def\be{\begin{eqnarray}&&} \def\ee{\end{eqnarray}}
\def \nonu {\nonumber \\&&}

\def\psla{\slash \! \! \!}
\begin{document}

\title{A light-front Quark Model for the 
electromagnetic  form factor of the pion}

\author{J. P. B. C. de Melo}
\address{IFT, Universidade Estadual Paulista, S\~ao Paulo, SP, Brazil}

\author{T. Frederico}
\address{ Dep. de F\'\i sica, ITA,
CTA, S\~ao Jos\'e dos Campos,
S\~ao Paulo, Brazil}

\author{E. Pace}

\address{Dipartimento di Fisica, Universit\`{a} di Roma "Tor Vergata" 
and
Istituto Nazionale di Fisica Nucleare, Sezione Tor Vergata, Via della Ricerca
Scientifica 1, I-00133 Roma, Italy}
\author{G. Salm\`e}

\address{Istituto Nazionale di Fisica Nucleare, 
Sezione Roma I, P.le A. Moro
2, I-00185 Roma, Italy}

\maketitle

\abstracts{In this contribution, 
an approach for a unified description of the pion electromagnetic 
form factor,  in the space- and time-like
 regions, within a constituent
quark model on the light front, will be reviewed. 
Our approach is based on  i)  the  
on-shell quark-hadron vertex functions in the valence sector,
 ii) the dressed photon vertex 
where a photon decays in a quark-antiquark pair, and iii) the emission and absorption amplitudes
 of a pion by a quark. Results  favorably compare with the existing
 experimental data.
}

\section{Introduction} Constituent quark models (CQM) developed within
a  light-front framework (for a general review, see, e.g., Ref. [\refcite{brodsky}])
appear to be an interesting tool for investigating the electromagnetic
properties of hadrons. One of the main advantage of the light-front approach is   
that the Fock vacuum (i.e. the vacuum of a free theory) could represent a 
reliable
approximation of the physical, interacting vacuum, given the positivity 
constraint and the
kinematical nature of the plus component of the total 
momentum, $P^+$. Then,  expanding hadron states over a  Fock 
basis becomes meaningful, and one can safely write for a meson
\be
| meson; P^+{\bf P}_{\perp};\zeta\rangle = \prod_{i=1,2}\int d{\bf k}_{i \perp}~d\xi_i
~\delta\left(\sum_{i=1,2}{\bf k}_{i \perp}\right )~
\delta\left(1-\sum_{i=1,2}\xi_i\right )~\times
\nonu \sum_{\{\zeta_j\}}\psi_{q\bar q}({\bf k}_{i \perp},\xi_i,\zeta_j)~
|q\bar{q};P^+ {\bf P}_{\perp};{\bf k}_{i \perp},\xi_i,\zeta_j\rangle +\nonu + 
\prod_{i=1,4}\int d{\bf k}_{i \perp}~d\xi_i
~\delta\left(\sum_{i=1,4}{\bf k}_{i \perp}\right )~
\delta\left(1-\sum_{i=1,4}\xi_i\right )~\times
\nonu \sum_{\{\zeta_j\}}\psi_{q\bar q q\bar q}({\bf k}_{i \perp},\xi_i,\zeta_j)~
|q \bar{q} q \bar{q};P^+ {\bf P}_{\perp};{\bf k}_{i \perp},\xi_i,\zeta_j\rangle +
\nonu +
~\prod_{i=1,3}\int d{\bf k}_{i \perp}~d\xi_i
~\delta\left(\sum_{i=1,3}{\bf k}_{i \perp}\right )~
\delta\left(1-\sum_{i=1,3}\xi_i\right )~\times
\nonu \sum_{\{\zeta_j\}}\psi_{q\bar q g}({\bf k}_{i \perp},\xi_i,\zeta_j)~|q \bar{q} ~g;P^+
 {\bf P}_{\perp};{\bf k}_{i \perp},\xi_i,\zeta_j\rangle +
 ..... 
\ee
where a light-front spin component along the $z$-axis is indicated by $\zeta$
  and the function $\psi$'s are the intrinsic amplitudes for the corresponding
Fock states. The possibility to include, in a given framework, states beyond the familiar
CQM ones (i.e. $|q\bar{q} \rangle$ for the mesons and $|qqq \rangle$ for the
baryons)   allows
one to address the rich phenomenology in the time-like (TL) region within a unified
approach, and, in turn, to increase the constraints to be fulfilled by a chosen 
model.

Our model\cite{plbpion} has been developed for investigating the electromagnetic form factor
 of a pion in the whole range of momentum transfer, i.e. for positive and
 negative squared mass  of the virtual photon. The first building block is
 represented by
  the Mandelstam formula for the matrix elements of the em current 
  \cite{mandel}. Following Ref. \refcite{mandel},
 the matrix elements for the pion in the TL region read
\be
j^{\mu}=e \frac{2 m^2}{\imath f^2_\pi} N_c\int
\frac{d^4k}{(2\pi)^4}
{\bar \Lambda}_{{\pi}}(k,P_{\pi})
 \Lambda_{\bar \pi}(k-P_{\pi},P_{{\bar \pi}}) ~\times \nonu  
Tr[S(k-P_{\pi}) \gamma^5
S(k-q) \Gamma^\mu S(k) \gamma^5 ]~~~~
\label{jmu}   
\ee 
where 
$
S(p)=[\psla{p}-m+\imath \epsilon]^{-1} \,$, 
with $m$ the mass of the constituent quark,
 $\Gamma^\mu(k,q)$ is the quark-photon vertex, $q^{\mu}$  the 
virtual-photon momentum, $ \Lambda_{{\pi}}(k,P_{{\pi}})$
  the pion vertex function, 
 $P^{\mu}_{\pi}$ and $P^{\mu}_{\bar \pi}$ are the pion and antipion momenta, respectively.
$N_c=3$ is the number of colors and the factor $2$ comes from the isospin
weight, since we are dealing with a charged pion form factor.
  For the space-like (SL) region, $P^{\mu}_{\pi}$
 has to be replaced by $-P^{\mu}_{\pi}$ and $\bar \pi$ by $\pi^{\prime}$.
 
A key mathematical step is given by the four dimensional integration of an
integrand that has, in principle, a very complicated analytical structure, due
to the presence of poles from the fermion propagators and from the 
analytical structure of the vertex
functions. A first approximation is introduced as 
 one projects out the Mandelstam formula on the light front by a $k^-$ 
 integration (see Figs 1 and 2). In particular,
 we have assumed that: i) the meson vertex functions
do not diverge in the complex plane $k^-$ for
$|k^-|\rightarrow\infty$ and  ii) the contributions of their singularities are negligible.
To emphasize the unified description of
the em form factor in TL and SL regions, the simplifying assumption of a chiral
pion ($m_\pi =0$) has been adopted. As a matter of fact, in the SL case we have 
carried out our analysis in a frame where
${\bf P}_{\pi \perp} = {\bf P}_{\pi \prime \perp} = {\bf 0} $,
obtaining that
${P}^+_{\pi} = q^+ (- 1 +
\sqrt{1- 4 m^2_\pi / q^2})/2 \ \ .$
In the limit $m_\pi \rightarrow 0$
one has
${P}^+_\pi=0 $ and ${P}^+_{\pi \prime} = q^+$. Then,
 only the contribution
of the pair-production mechanism survives, see Fig. 1(b).

In the TL case,
   the  choice
${\bf P}_{\bar{\pi} \perp} = -{\bf P}_{{\pi} \perp}={\bf 0} $ leads to
$P^+_{{\pi}} = q^+ (1 \pm \sqrt{1-4 m^2_\pi/ q^2})/2.$
In the limit $m_\pi \rightarrow 0$,
 in analogy with the SL case,  we have adopted the choice 
 $P^+_\pi=0$. Then, only the contribution of the
diagram (b) in Fig. 2 survives.
\begin{figure}
 \begin{center}
\includegraphics[width=10.cm]{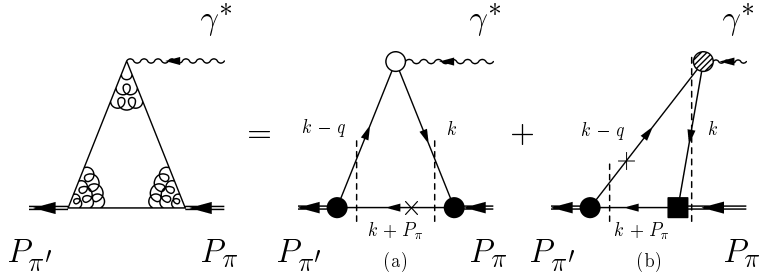}
\end{center}
\caption{Light-front time-ordered diagrams contributing to the em form factor of the pion in the space-like
region. In contribution $(a)$ one has $0<k^+ + P^+_\pi<P^{ +}_\pi$, while in $(b)$ $0<k^+
< q^+$. Vertical dashed lines indicate a fixed value for the light-front time
$x^+$ that flows from the right hand toward the left hand.}
\footnotesize{After Ref. [\refcite{plbpion}]}
\end{figure}
\begin{figure}
\begin{center}
\includegraphics[width=10.cm]{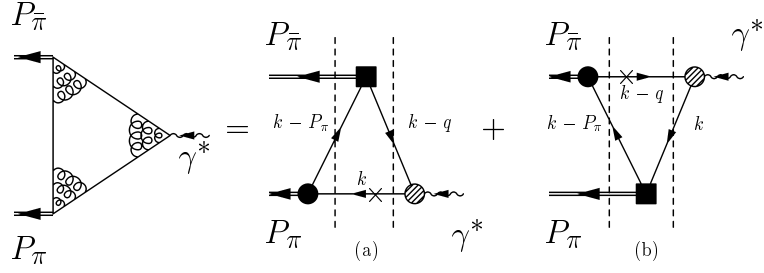}
\end{center}
\caption{Light-front time-ordered diagrams contributing to the em form factor of the pion in the time-like
region. In contribution $(a)$ one has $0<k^+ < P^+_\pi$, while in $(b)$ $P^+_\pi<k^+
< q^+$.}
\footnotesize{After Ref. [\refcite{plbpion}]}\end{figure}
To complete our model\cite{plbpion} we have to answer the
 following questions : i) how to describe  the
$q\bar{q}$-meson vertices ? ii) how to model the dressed quark-photon 
vertex ?  
  iii) how to deal with  the amplitude for the
emission or absorption of a pion by a quark ? In the following section, our answers will
be illustrated in some detail. 
 
 \section{The Model} In order to construct our model,  we have first introduced
 an approximation 
 for the momentum component of the pion vertex function, $\Lambda_\pi
 (k,P_\pi)$,
  when both the quarks are on their mass shell
 and in the interval  $0\leq k^+ \leq P^+_\pi$. Such a
 function of the quark light-front momenta,
 can be approximated by the "light-front pion wave function", $\psi_{\pi}$, obtained
 within the Hamiltonian light-front dynamics, 
 through the following equation
 \be(\psla k_{on} + m) \gamma^5 
  \psi_{\pi}(k^+, {\bf k}_{\perp}; P^+_{\pi}, {\bf P}_{\pi \perp})
\left [(\psla k - \psla P_{\pi})_{on} + m \right ] =
\nonu
(\psla k_{on} + m) \gamma^5 ~ \frac{m}{f_\pi}
~\frac{P^+_{\pi}}{[m^2_\pi - M^2_0(k^+, {\bf k}_{\perp}; P^+_{\pi},
{\bf P}_{\pi \perp})]} ~ [ \Lambda_{\pi}(k,P_{\pi}) ]_{[k^- = k^-_{on}]}~\times
\nonu
\left [(\psla k - \psla P_{\pi})_{on} + m \right ]
\label{wfp}
\ee
 where $k^-_{on}= ({\bf k}_{\perp}^{2}+m^2)/k^+$ and $M_0$ is the light-front 
 free mass (see, e.g., Ref. [\refcite{Jaus90}]).
A similar relation is also adopted for the vector mesons (VM). 
 For the pion and the VM wave functions,  we have used the
eigenfunctions of the square 
mass operator proposed in Refs. [\refcite{tobpauli,FPZ02}],  within a relativistic  
constituent quark model which achieves a  natural 
explanation of the "Iachello-Anisovitch law"~\cite{Iach,ani}. 
The VM eigenfunctions
, $\psi_{n}(k^+, {\bf k}_{\perp}; q^+,{\bf q}_{\perp})$,  
are normalized to the probability 
of the lowest ($q\bar q$) Fock state, roughly estimated to be $\sim 1/\sqrt{2n + 3/2}$ in a 
simple model \cite{plbpion} 
that reproduces the "Iachello-Anisovitch law" \cite{Iach,ani}($n$ is the
principal quantum number).

As for  the second question, 
following Ref. [\refcite{plbpion}], a  Vector Meson Dominance (VMD) approximation
 is applied to the
quark-photon vertex $\Gamma^\mu(k,q)$, when a $q\bar{q}$ pair is produced. In
particular,  the
plus component of the quark-photon vertex  reads as follows (see Fig. 3)
  \be
\Gamma^+(k,q) =  \sum_{n, \lambda}~
\left [ \epsilon_{\lambda} \cdot \widehat{V}_{n}(k,k-q)  \right ]   
\Lambda_{n}(k,P_n) ~\times \nonu
{ \sqrt{2} [\epsilon ^{+}_{\lambda}]^* f_{Vn} \over (q^2 -
M^2_n + \imath M_n \Gamma_n(q^2))}
\label{vert}  
\ee
where $f_{Vn}$ is the decay constant of the n-th vector
meson into a virtual photon, 
 $M_n$($P_n$)  the mass (four-momentum) of the VM, 
 $\Gamma_n(q^2)=\Gamma_n q^2/M^2_n$ (for $q^2>0$) 
 the corresponding total decay width,
 $\epsilon_{\lambda}$  the VM polarization, and    
  $ [ \epsilon_{\lambda} \cdot \widehat{V}_{n}(k,k-q)    ~ 
\Lambda_{n}(k,q)$  the VM vertex function ( 
for the  VM Dirac structure, $\widehat{V}_{n}$, see below).
\begin{figure}
 \begin{center}
\includegraphics[width=8.cm]{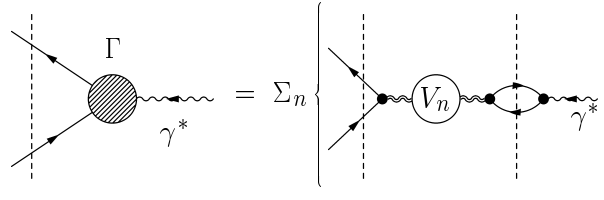}\end{center}
\caption{Diagrammatic analysis of the quark-photon vertex, $\Gamma^\mu(k,q)$.}
\footnotesize{After Ref. [\refcite{plbpion}]}
\end{figure}
The decay constant, $f_{Vn}$, 
is evaluated 
starting from a four dimensional representation 
in terms of the VM Bethe-Salpeter vertex and integrating over $k^-$.
In particular, one has
\be 
f_{Vn} = -\imath \frac{N_c} {4 (2\pi)^4} ~ 
\int dk^- dk^+ d{\bf k}_{\perp} 
\frac{Tr[ \gamma^+ ~ {\mathcal {V}}_{nz}(k,k-P_n)] ~ \Lambda_{n}(k,P_n)} 
{[k^2 - m^2 + \imath \epsilon] [(P_n - k)^2 - m^2 + \imath \epsilon]} ~ =    
\nonu
= - {N_c \over 4 (2\pi)^3} \int_0^{P_n^+} dk^+ ~ d{\bf k}_{\perp} 
{ Tr \left [  \gamma^+  ~  {\mathcal V}_{nz}(k,k-P_n) \right ]
\over k^+ ~ (P_n^+ - k^+)} ~ \psi_{n}(k^+, {\bf k}_{\perp}; M_{n}, {\vec 0}_{\perp})  
\label{fV1}  
\ee  
where ${\mathcal V}_{nz}(k,k-P_n) = (\psla k - \psla P_n + m)  
\widehat{V}_{n z}(k,k-P_n) (\psla k + m)$. 

Note that only $^3S_1$ isovector vector mesons
 have been taken into account in the calculations, and therefore  
 the on-mass-shell spinorial part of the VM vertex must reproduce 
  the well-known Melosh rotations for $^3S_1$ 
states, (see, e.g., Ref. [\refcite{Jaus90}]), viz.
\be 
\widehat{V}^{\mu}_{n}(k,k-q) = \gamma^{\mu} - 
{k^{\mu}_{on}-(q-k)_{on}^{\mu} \over M_{0}  
(k^+, {\bf k}_{\perp}; q^+, {\bf q}_{\perp}) + 2 m } ~~\ . 
\label{gams1}  
\ee
 The third question raised in the Introduction is answered
 by describing with a constant\cite{JI01} the amplitude for the
emission or absorption of a pion by a quark, i.e. 
  the pion vertex function in the non-valence 
sector. 
The value of the constant is fixed by the pion charge normalization, given our
simplifying assumption of a chiral pion. 
 
 In the limit $m_{\pi} \rightarrow 0$ the pion form factor 
 receives contributions only from processes
where the photon decays in a $q\overline q$ pair. 
Then, by means of Eq. (\ref{vert}) the matrix element $j^+$ can be written as a sum  
 over the vector mesons, and 
 consequently the form factor becomes 
 \be 
 F_{\pi}(q^2) = \sum_n~ {f_{Vn} \over q^2 - M^2_n + \imath M_n \Gamma_n(q^2)} ~ 
 g^+_{Vn}(q^2) 
\label{tlff}  
\ee 
where $g^+_{Vn}(q^2)$, for $q^2 > 0$, is the form factor for the VM decay in a pair of pions. 
 
Each VM contribution to the sum (\ref{tlff}) is 
 invariant under kinematical light-front boosts and 
 can be evaluated in the rest frame of the  
corresponding resonance (with $q^+=M_n$ and ${\bf q}_{\perp}=0$). 

  It turns out that the same expression for $g^+_{Vn}(q^2)$ holds
 both in the TL and in the SL  
 regions\cite{plbpion}.

\section{Results}
In our calculations the up-down quark mass is fixed at 
0.265 $GeV$ \cite{FPZ02} and the oscillator strength at $\omega = 1.39~GeV^2$  \cite{ani}. 
For the first four vector mesons  
the known experimental masses and widths are used \cite{pdg} (see Table I), a
part the value of the mass of $\rho$-meson changed to  $m_\rho=0.750~Gev$, in
order to reproduce the correct position of the $\rho$ peak in the time-like 
form
factor of the charged pion. 
For the  VM with $M_n >~2.150~GeV$, the mass values corresponding to the model 
of Ref. \refcite{FPZ02} 
are used, while for the unknown widths we use a single value 
$\Gamma_n = 0.15 ~GeV$. To obtain  stability of the results up to $q^2 =~10 ~(GeV/c)^2$
twenty resonances are considered.
\begin{table}
\begin{center}\quote{ Table I. Input values of our model, $M_n(PDG)$ and 
$\Gamma_{n}(PDG)$ (see text). For the sake of comparison, the masses
evaluated in the model of Refs. [\refcite{tobpauli,FPZ02}], $M_n$(FPZ) are also
shown (the mass of the $\rho$ is an input for this model).}
\end{center} \begin{center}
\begin{tabular} {|c||c|c| |c| }
\hline
 VM  & $M_n$(FPZ) & $M_n(PDG)$  & $\Gamma_{n}(PDG)$ \\
\hline
$\rho(770)$  & \em{.770} GeV & {.770 GeV} & {0.15 GeV }\\
\hline
$\rho(1450)$ &1.408 GeV & {1.465 GeV} & {0.37 GeV} \\
\hline
$\rho(1770)$ & 1.836 GeV & {1.723 GeV} & {0.30 GeV }\\
\hline
$\rho(2150)$ & 2.182 GeV & {2.149 GeV} & {0.18 GeV} \\
\hline
\end{tabular}
\end{center}
\end{table}

 The calculated pion form factor is shown in Fig. 4 
in a wide region of square 
momentum transfers, from $-10$ $(GeV/c)^2$ up to 10 $(GeV/c)^2$. The VM dominance ansatz 
for the (dressed photon )-($q\overline q$)
vertex, within a CQ model consistent with the meson spectrum, 
is able to give a unified description of the pion form factor both in the SL and
TL regions.

\begin{figure}
\vspace{0.5cm}
\includegraphics[width=7.9cm,angle=-90]{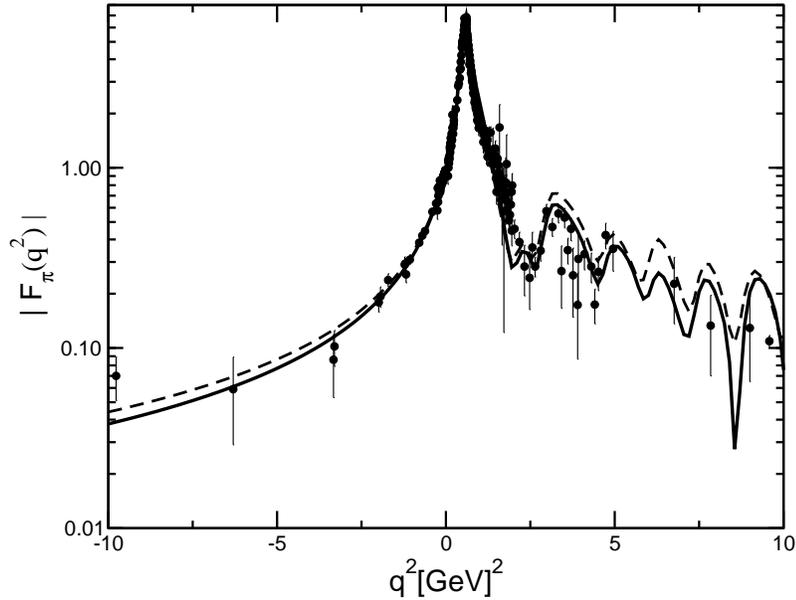}

$~~~~~~~~~~~~~~~~ $
\caption{Pion electromagnetic form factor vs. the square momentum
transfer $q^2$. Solid line: result obtained by using the full
pion wave function.  Dashed line: result obtained}\footnotesize{ by using the asymptotic 
pion wave function (After Ref.
\refcite{plbpion}).}
\end{figure}
The SL form factor is notably 
well described, see Fig. 5, from the high-$q^2$ region to the low-$q^2$ one, as well
as the charge radius. 
Finally, it is worth noting that the heights of the TL bumps directly depend 
on the calculated values of $f_{Vn}$ and $g^+_{Vn}$, and for the sake of
completeness a table of calculated em decay widths, $\Gamma^{th}_{e^+e^-}$, 
are shown in Table II.
\begin{figure}
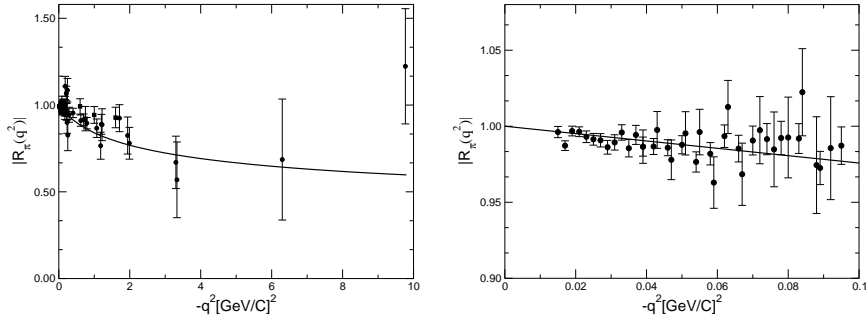

\parbox{5.5cm}{\includegraphics[width=5.5cm]{ratiosl.eps}} $~~~$
\parbox{5.5cm}{\includegraphics[width=5.5cm]{ratiozoom.eps}}
\caption{The ratio $R_\pi(q^2)=
F_\pi(q^2)/\left [ 1/(1-q^2/m^2_\rho)\right ]$ vs $q^2$,
in the SL region ($m_\rho=0.770~GeV$). The rightmost figure shows in great detail the region 
$0~\le ~-q^2 ~\le~ 0.1 (GeV/c)^2$, relevant for evaluating the charge radius.}\footnotesize{ Experimental data from R. Baldini  
et al.\cite{baldini}}   
\end{figure}

\begin{table}
\begin{center}
\quote{Table II. Em decay widths, $\Gamma^{th}_{e^+e^-}$, for the first three
$\rho$-mesons}
\end{center} \begin{center}
\begin{tabular} {|c||c|c|}
\hline
 VM  & $\Gamma^{th}_{e^+e^-}$  & $\Gamma^{exp}_{e^+e^-}$ \\
\hline
$\rho(770)$  & 6.37 KeV & 6.77 $\pm$ 0.32 KeV \\
\hline
$\rho(1450)$  & 1.61 KeV & $>$ 2.30 $\pm$ 0.50 KeV \\
\hline
$\rho(1770)$  & 1.23 KeV & $>$ 0.18 $\pm$ 0.10 KeV \\
\hline
\end{tabular}
\end{center}
\end{table}

\section{Perspectives}
 The  results obtained within our approach\cite{plbpion} encourage 
 an investigation of the TL form factors of the nucleon
   based on  a simple ansatz for the non-valence component
 of the nucleon state, following as a  guideline  the pion case.

\end{document}